\documentstyle[11pt,aaspp4,times]{article}
\makeatletter
\let\@internalcite\cite
\def\cite{\def\@citeseppen{-1000}%
    \def\@cite##1##2{(##1\if@tempswa , ##2\fi)}%
    \def\citeauthoryear##1##2##3{##1 ##3}\@internalcite}

\def\citeNP{\def\@citeseppen{-1000}%
    \def\@cite##1##2{##1\if@tempswa , ##2\fi}%
    \def\citeauthoryear##1##2##3{##1 ##3}\@internalcite}
\def\citeN{\def\@citeseppen{-1000}%
    \def\@cite##1##2{##1\if@tempswa , ##2)\else{)}\fi}%
    \def\citeauthoryear##1##2##3{##1 (##3}\@citedata}
\def\citeA{\def\@citeseppen{-1000}%
    \def\@cite##1##2{(##1\if@tempswa , ##2\fi)}%
    \def\citeauthoryear##1##2##3{##1}\@internalcite}
\def\citeANP{\def\@citeseppen{-1000}%
    \def\@cite##1##2{##1\if@tempswa , ##2\fi}%
    \def\citeauthoryear##1##2##3{##1}\@internalcite}
\def\shortcite{\def\@citeseppen{-1000}%
    \def\@cite##1##2{(##1\if@tempswa , ##2\fi)}%
    \def\citeauthoryear##1##2##3{##2 ##3}\@internalcite}
\def\shortciteNP{\def\@citeseppen{-1000}%
    \def\@cite##1##2{##1\if@tempswa , ##2\fi}%
    \def\citeauthoryear##1##2##3{##2 ##3}\@internalcite}
\def\shortciteN{\def\@citeseppen{-1000}%
    \def\@cite##1##2{##1\if@tempswa , ##2)\else{)}\fi}%
    \def\citeauthoryear##1##2##3{##2 (##3}\@citedata}
\def\shortciteA{\def\@citeseppen{-1000}%
    \def\@cite##1##2{(##1\if@tempswa , ##2\fi)}%
    \def\citeauthoryear##1##2##3{##2}\@internalcite}
\def\shortciteANP{\def\@citeseppen{-1000}%
    \def\@cite##1##2{##1\if@tempswa , ##2\fi}%
    \def\citeauthoryear##1##2##3{##2}\@internalcite}
\def\citeyear{\def\@citeseppen{-1000}%
    \def\@cite##1##2{(##1\if@tempswa , ##2\fi)}%
    \def\citeauthoryear##1##2##3{##3}\@citedata}
\def\citeyearNP{\def\@citeseppen{-1000}%
    \def\@cite##1##2{##1\if@tempswa , ##2\fi}%
    \def\citeauthoryear##1##2##3{##3}\@citedata}

\def\@citedata{%
	\@ifnextchar [{\@tempswatrue\@citedatax}%
				  {\@tempswafalse\@citedatax[]}%
}

\def\@citedatax[#1]#2{%
\if@filesw\immediate\write\@auxout{\string\citation{#2}}\fi%
  \def\@citea{}\@cite{\@for\@citeb:=#2\do%
    {\@citea\def\@citea{, }\@ifundefined
       {b@\@citeb}{{\bf ?}%
       \@warning{Citation `\@citeb' on page \thepage \space undefined}}%
{\csname b@\@citeb\endcsname}}}{#1}}%

\def\@citex[#1]#2{%
\if@filesw\immediate\write\@auxout{\string\citation{#2}}\fi%
  \def\@citea{}\@cite{\@for\@citeb:=#2\do%
    {\@citea\def\@citea{; }\@ifundefined
       {b@\@citeb}{{\bf ?}%
       \@warning{Citation `\@citeb' on page \thepage \space undefined}}%
{\csname b@\@citeb\endcsname}}}{#1}}%

\def\@biblabel#1{}

\newlength{\bibhang}
\makeatother

\def\LA{Ly$\alpha$}

\def\ebv{{\rm E}(\bv)}

\def\etal{{et al.}}
\newcommand{\fig}[1]{Fig.~\ref{#1}}

\def\hst{{\it HST}}

\def\ie{i.e.}
\def\iue{{\it IUE}}
\def\mh{{\rm [M/H]}}

\begin{document}

\title{Constraints on the Horizontal-Branch Morphology\\
 of the Globular Cluster M79 (NGC 1904)\\
 from Optical and Far-UV Observations}


\author{W. Van Dyke Dixon and Arthur F. Davidsen}
\affil{Department of Physics and Astronomy, The Johns Hopkins University\\
3400 N. Charles Street, Baltimore, Maryland 21218\\
wvd@pha.jhu.edu, afd@pha.jhu.edu}

\author{Ben Dorman}
\affil{Laboratory for Astronomy and Solar Physics, Code 681 \\
NASA Goddard Space Flight Center, Greenbelt, MD 20771 \\
dorman@shemesh.gsfc.nasa.gov}

\and

\author{Henry C. Ferguson}
\affil{Space Telescope Science Institute, Baltimore, MD 21218 \\
ferguson@stsci.edu}

\begin{center}{Version of \today}\end{center}

\begin{abstract}

The globular cluster M79 was observed with the Hopkins Ultraviolet
Telescope (HUT) during the Astro-1 space shuttle mission in 1990
December. The cluster's far-UV integrated spectrum shows strong
absorption in the Lyman lines of atomic hydrogen. We seek to use this
spectrum, together with optical photometry, to constrain the stellar
mass distribution along its zero-age horizontal branch (ZAHB). We find
that a Gaussian distribution of ZAHB masses, with a mean of $0.59
M_{\sun}$ and standard deviation $0.05 M_{\sun}$, is able to reproduce
the cluster's $(B,V)$ color-magnitude diagram when subsequent stellar
evolution is taken into account, but cannot reproduce the cluster's
far-UV spectrum.  Model stellar spectra fit directly to the HUT data
indicate a surprising distribution of atmospheric parameters, with
surface gravities (and thus implied masses) significantly lower than
are predicted by canonical HB evolutionary models. This result is
consistent with the findings of Moehler et al. [A\&A, 294, 65 (1995)]
for individual HB stars in M15. Further progress in understanding the
mass distribution of the HB must await resolution of the
inconsistencies between the derived stellar atmospheric parameters and
the predictions of HB evolutionary models.  Improved stellar spectral
models, with higher spectral resolution and non-solar abundance ratios,
may prove useful in this endeavor.

\end{abstract}


\section{Introduction}
 
M79 (NGC 1904) is a centrally condensed, intermediate-metallicity
([Fe/H] = $-1.69$) globular cluster \shortcite{DK86,Djorgovski93}
with an extremely blue horizontal branch \shortcite{SH77}, extending at its
high-temperature end as faint as the main-sequence turnoff
\shortcite{Ferraro92}. The cluster's integrated UV spectrum is nearly flat
from 1500 to 3300 \AA\ \shortcite{vAlbada81}, and observations with the
Ultraviolet Imaging Telescope (UIT; \shortciteNP{UIT_M79}) indicate that
most of the far-UV flux emanates from individual blue HB (BHB) and post-HB
stars.

In our current understanding \shortcite{Renzini81p319},
stars with zero-age-main-sequence mass $M_{ZAMS} \sim M_{\sun}$
lose about $0.2 M_{\sun}$ while ascending the red-giant branch (RGB)
and eventually produce zero-age horizontal branch (ZAHB) stars with core
masses $M_C \sim 0.5 M_{\sun}$ and envelope masses $M_{env} \sim
0.1 M_{\sun}$. A star's location on the ZAHB is determined by the mass
of the hydrogen-rich envelope remaining above the helium-rich core,
which for typical globular cluster ages is a function only of the
abundances of helium and heavier elements $(Y,Z)$.
Stellar evolutionary models imply a spread in
$M_{env}$, with the bluest HB stars having the lowest masses
\shortcite{Dorman93}.

The mass distributions used to derive gross cluster properties from the
HB population are most often based on the function introduced by
\shortciteN{Rood73}. This function consists of a Gaussian multiplied by
polynomial truncation terms designed to ensure that the simulation
contains physically possible objects, \ie, stars with $M_C \le M \le
M_{\rm RG}$. Otherwise, the assumption is not based on any physical
understanding of the cause of the HB mass spread \shortcite{Rood90p11}.
Monte-Carlo simulations based on such syntheses have been quite
successful in reproducing observed HB morphologies
(\shortciteNP{Rood73}; \shortciteNP{LDZ90}, 1994; \shortciteNP{Cat93});
however, the polynomial truncation terms give distributions that are
skewed away from the theoretical limits of the mass range, a situation
that may not be desirable for the extremely blue HB of M79.

In this paper we analyze the far-UV spectrum of M79 obtained with the
Hopkins Ultraviolet Telescope (HUT). Our object is to derive
constraints on the ZAHB mass distribution from our far-UV spectrum and
from optical photometry. \shortciteN{Ferraro92} note that the
distribution of stars in the color-magnitude diagram (CMD) of M79 seems
quite different from a Gaussian. We investigate whether a Gaussian 
distribution of masses along the ZAHB
can reproduce this asymmetry by combining the HB
evolutionary models of \shortciteN{Dorman93} with the synthetic stellar
flux distributions of \shortciteN{Kurucz92} to fit the cluster's optical CMD
\shortcite{Ferraro92}. We find that a Gaussian distribution of ZAHB masses
with a mean of $0.59 M_{\sun}$ and standard deviation $0.05 M_{\sun}$
can indeed reproduce the cluster's CMD. We use this model to compute
synthetic far-UV spectra using a Monte Carlo algorithm, compare them
with the HUT spectrum, and find that it cannot reproduce the observed
far-UV flux distribution. Using individual Kurucz model flux
distributions, we derive surface gravities for BHB stars that are
significantly lower than theoretical predictions.  This result is
consistent with that of \shortciteN{MHd95} for individual HB
stars in M15. We conclude that more precise constraints on the HB mass
distribution must await a resolution of this discrepancy.

\section{Observations and Data Reduction}

\begin{figure}[t]
\plottwo{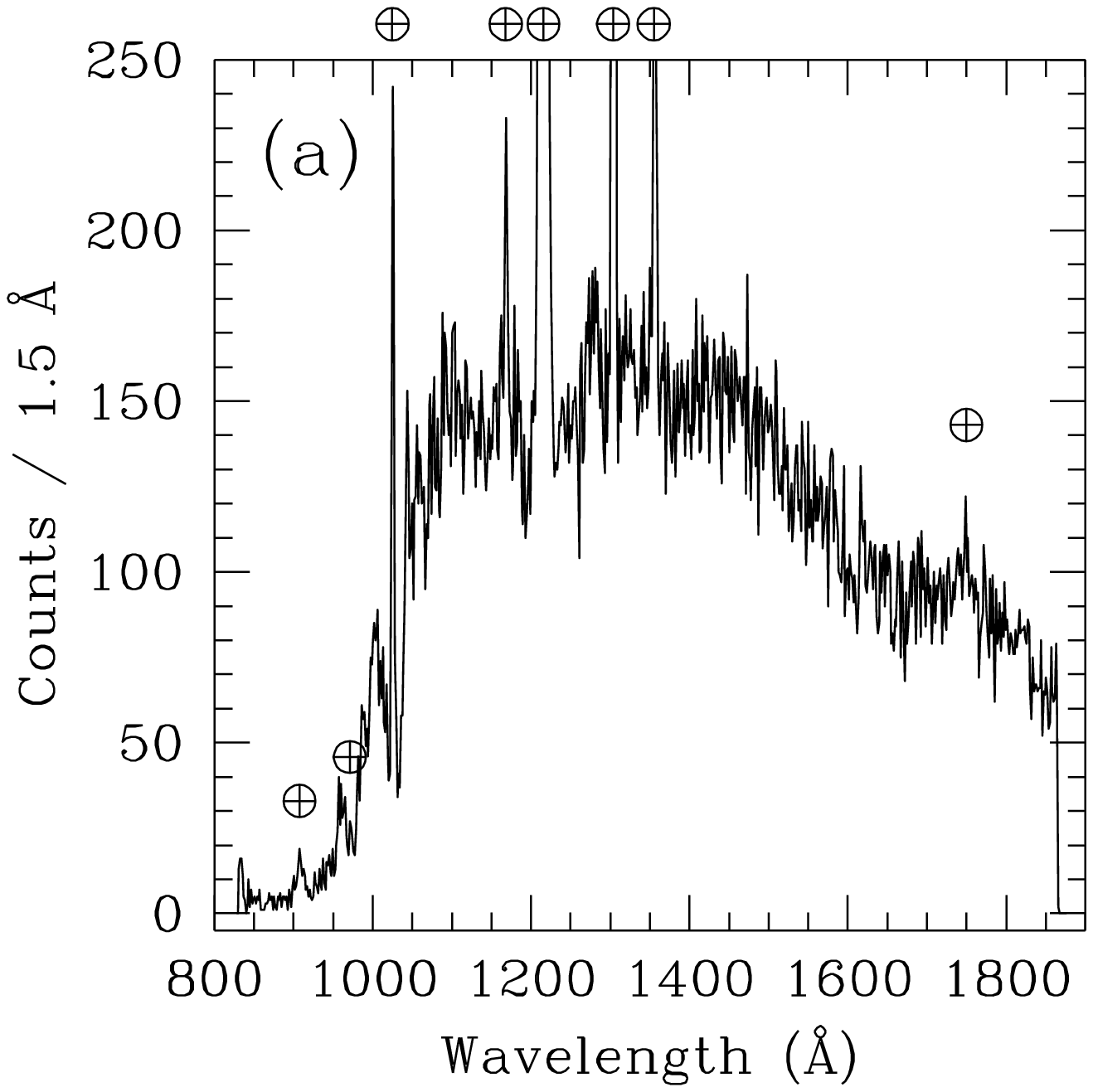}{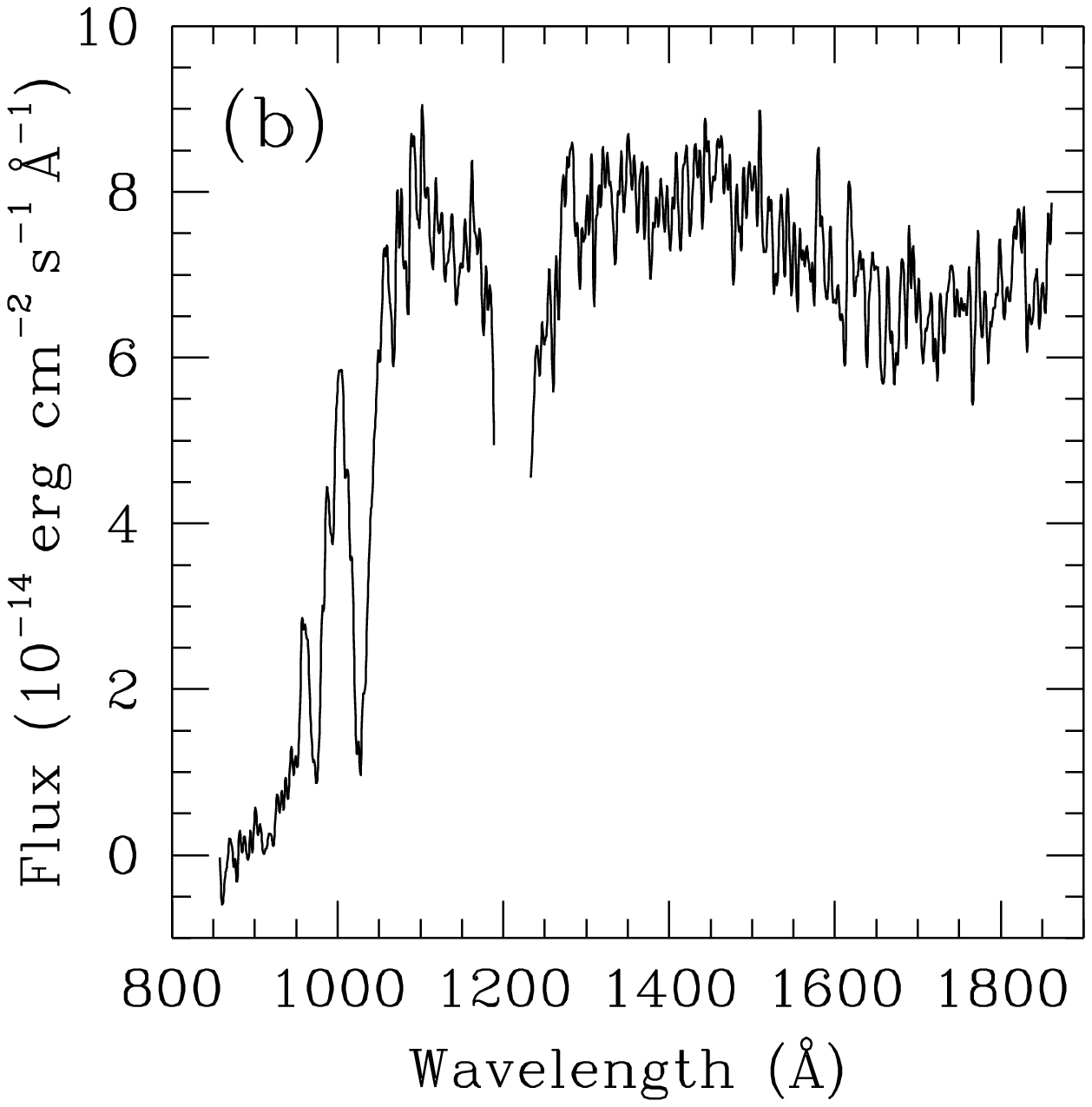}
\caption
{(a) HUT spectrum of M79 in raw counts, binned by three pixels
(about 1.5 \AA) and uncorrected for airglow or extinction. Airglow
features are marked with \earth. (b) Flux-calibrated,
airglow-subtracted spectrum of M79. The spectrum has been dereddened
with a Cardelli \etal\ (1989) extinction curve, assuming \ebv\ = 0.01
and $R_V = 3.1$, and smoothed with a Gaussian to 3.0 \AA\ resolution.
\label {HUT_RAW} 
}
\end{figure}

The spectrum of M79 was obtained with HUT on the Astro-1 mission of the
space shuttle {\it Columbia} in 1990 December. HUT consists of a 0.9-m
mirror that feeds a prime-focus spectrograph with a microchannel-plate
intensifier and photo-diode array detector. First-order sensitivity
covers the region from 830 to 1850 \AA\ at 0.51 \AA\ ${\rm pixel}^{-1}$
with about 3 \AA\ resolution. The spectrograph and telescope are
described in detail by \shortciteN{HUT_INSTR}. Two observations of
M79 were obtained during orbital night through a $9\arcsec \times
116\arcsec$ aperture, with a total integration time of 2366~s.  The
first observation, of 764~s duration, was begun on 1990 December 5 at
UT 15:33:47; the second, for 1602~s, was obtained two orbits later,
beginning at UT 18:33:45.  During both observations, the aperture was
centered on the cluster center of light. The pointing stability was
quite good: 1.0\arcsec\ rms in both pitch and yaw for the 1602 s
observation and 1.4\arcsec\ and 1.8\arcsec\ rms, respectively, in pitch
and yaw for the 764 s pointing.  (Pitch corresponds to the short
dimension of the HUT aperture, yaw to the long dimension.) The combined
spectrum, presented in \fig{HUT_RAW}a, shows strong absorption by
atomic hydrogen, but no emission features other than well-known
geocoronal lines.

In order to estimate the contribution of the earth's residual
atmosphere to the observed spectrum, emission-line profiles were fit to
an airglow spectrum taken during orbital night through the same
aperture as the M79 observations. These airglow line profiles were
combined with a linear continuum and fit to the region around each
airglow line in the cluster spectrum. The line profiles were held fixed
while their fluxes were allowed to vary. Exception was made for the
Ly$\beta$ and $\gamma$ airglow lines, for which a Gaussian
absorption feature was included in the continuum-plus-airglow model
that was fit to the data, and the complex of \ion{N}{1} and \ion{N}{2}
airglow features between 903 and 916 \AA, for which the model
line shapes, as well as the fluxes, were allowed to vary in fitting the
cluster spectrum. The resulting synthetic airglow spectrum was
subtracted from the cluster spectrum. This procedure led to an
oversubtraction of the core of Ly$\alpha$; regions around this line
were omitted from subsequent model fits to the spectra. The airglow
lines subtracted are marked in \fig{HUT_RAW}a.
 
Once airglow-subtracted, the spectrum was scaled to correct for
pulse-persistence effects. A constant dark-current contribution was
subtracted, as was scattered light from the grating, determined
from the count rate between 850 and 895 \AA, a region free from
airglow and stellar emission. The second-order contribution from
wavelengths longward of 912 \AA, which affects fluxes longward of 1824 \AA,
was also scaled and subtracted. The raw-counts spectrum was converted
to physical units using the HUT absolute calibration, based on a model
atmosphere of the white dwarf G191-B2B. Comparison with pre-flight and
post-flight laboratory measurements indicates that the calibration is
accurate to approximately 5\% over all wavelengths
\shortcite{HUT_INSTR,Kimble93,HUT1CAL2}.
Errors for each channel were estimated assuming Gaussian
statistics and were propagated through the data-reduction process.
The flux-calibrated, airglow-subtracted HUT spectrum of M79 is
presented in \fig{HUT_RAW}b. For this figure (only), the
spectrum has been dereddened with a \shortciteN{CCM89} curve,
assuming \ebv\ = 0.01 and $R_V = 3.1$, and smoothed with a Gaussian
to 3.0 \AA\ resolution.

\begin{figure}[t]
\plottwo{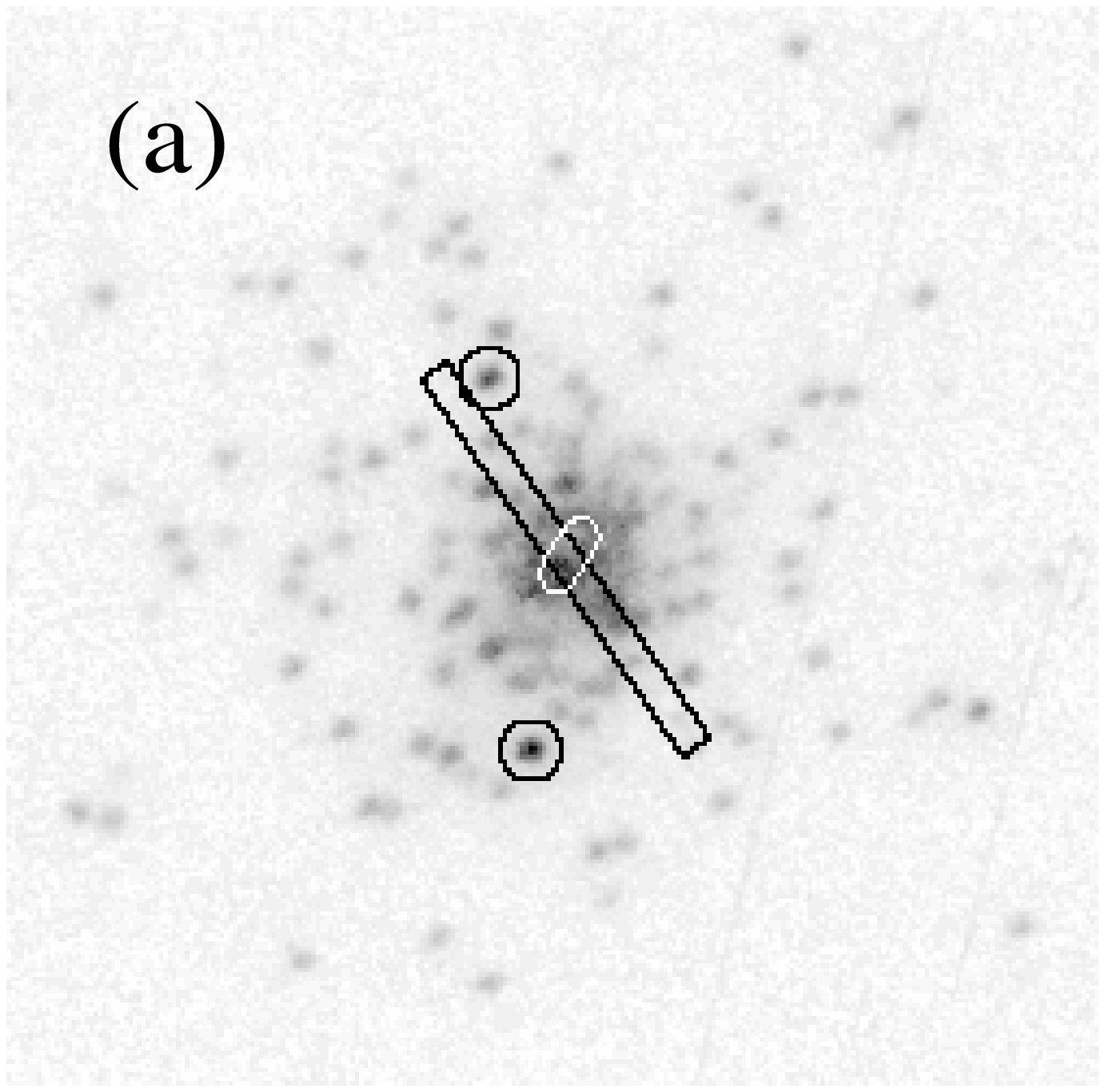}{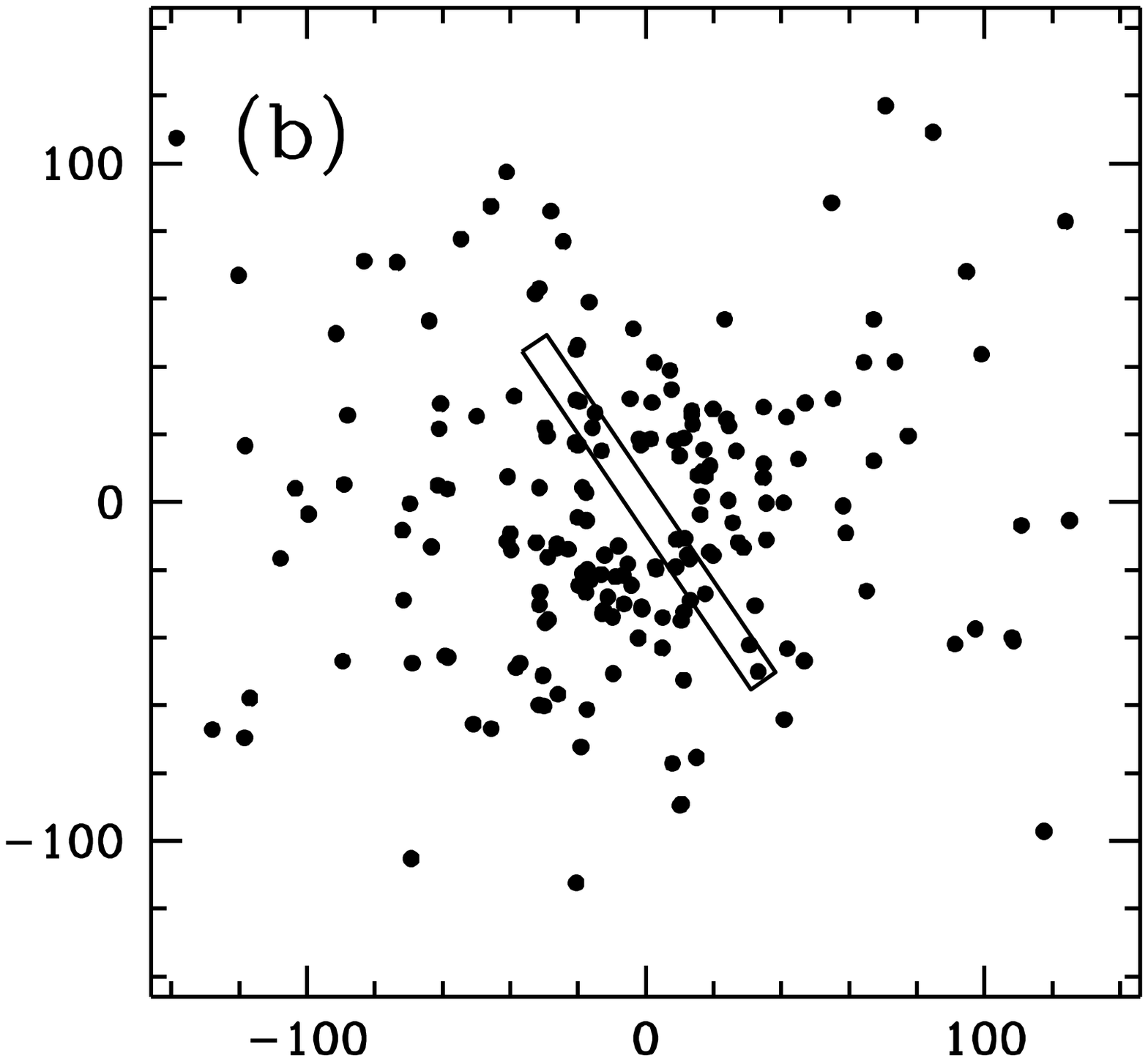}
\caption
{(a) Far-UV image of M79 obtained with the Ultraviolet Imaging
Telescope (UIT; Hill \etal\ 1992). North is up and east is to the
left. The position of the HUT aperture is marked in black, and that of
the $IUE$ aperture (Altner \& Matilsky 1993) in white. For scale, the
length of the HUT aperture is 116\arcsec. Two UV-bright stars
discovered by Hill \etal\ are circled. The UV-bright star
discovered by Altner \& Matilsky lies near the southeastern
intersection of the HUT and $IUE$ apertures. (b) Map showing the
positions of HB branch stars of M79 drawn from the data of Ferraro
\etal\ (1992). Units are arc seconds from the cluster center.
\label {UIT}
}
\end{figure}

The exact position of the HUT aperture is shown in \fig{UIT}a, a
far-UV image of the cluster obtained by UIT \shortcite{UIT_M79}. Two
UV-bright stars discovered by UIT are circled. \shortciteANP{UIT_M79}
estimate that M79 contains a total of $220 \pm 10$ HB stars. From
\fig{UIT}a, we estimate that 20 to 40 of these stars fall within the
HUT aperture. Also shown is the location of the {\it International
Ultraviolet Explorer's} ($IUE$) large aperture during an observation
made by \shortciteN{Altner93}. Their analysis indicates the presence
of a UV-bright star at the edge of the HUT aperture, near the
southeastern intersection of the HUT and $IUE$ apertures. A second
UV-bright star found by \shortciteANP{Altner93} is much fainter and
lies outside the HUT aperture. These objects were not known at the time
of the Astro-1 flight.

\begin{figure}[t]
\plottwo{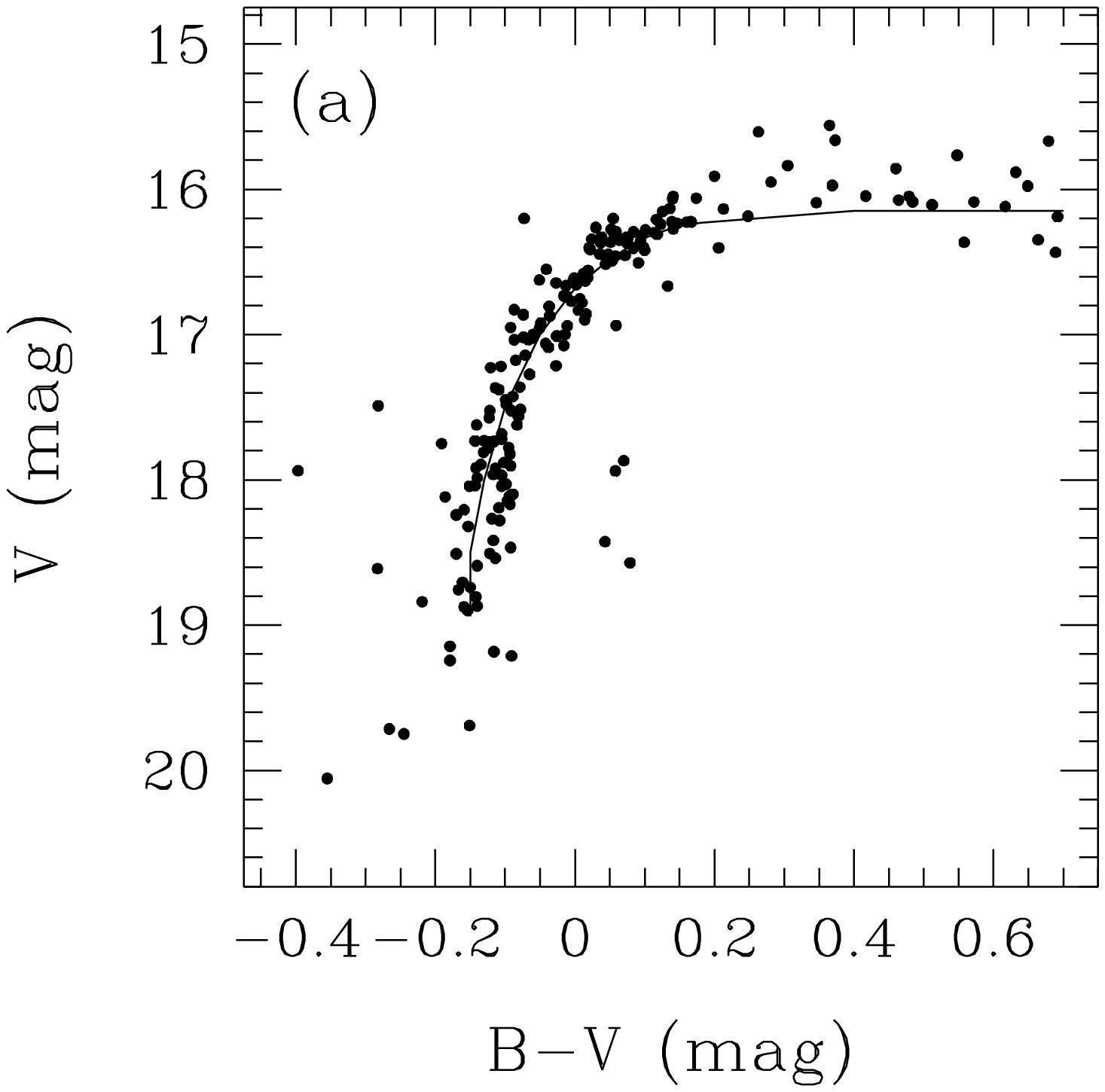}{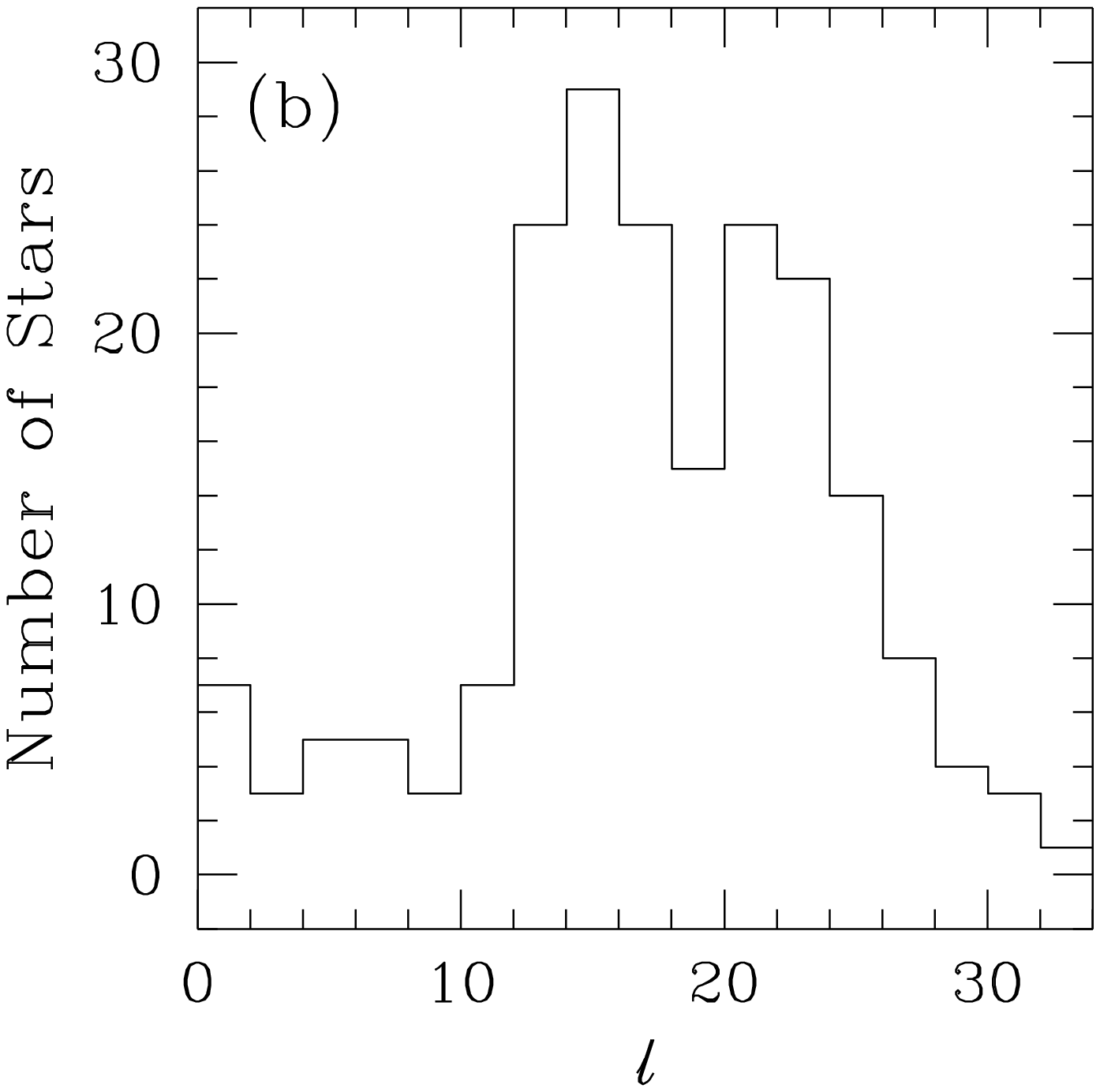}
\caption
{(a) Color-magnitude diagram (CMD) of the approximately 200 HB stars observed
by Ferraro \etal\ (1992). The ridge line for the cluster is also
shown. (b) The observed distribution of stars along the HB in M79. 
$l \equiv 0$ at $V = 16.15$ and $\bv = 0.70$ and increases to the blue
along the ridge line.
\label {M79} 
}
\end{figure}

\shortciteN{Ferraro92} provide positional information, colors, and
magnitudes for stars farther than 15\arcsec\ from the center of M79. A
map based on their positions of the cluster's HB stars is presented in
\fig{UIT}b. Note the correspondence between this map and the UIT image;
most of the stars resolved with UIT were observed by
\shortciteANP{Ferraro92} A CMD for the complete sample of approximately
200 HB stars in M79, statistically corrected for contamination by field
stars, is presented in \fig{M79}a.

\section{Fitting the Optical CMD}

\begin{figure}[t]
\plottwo{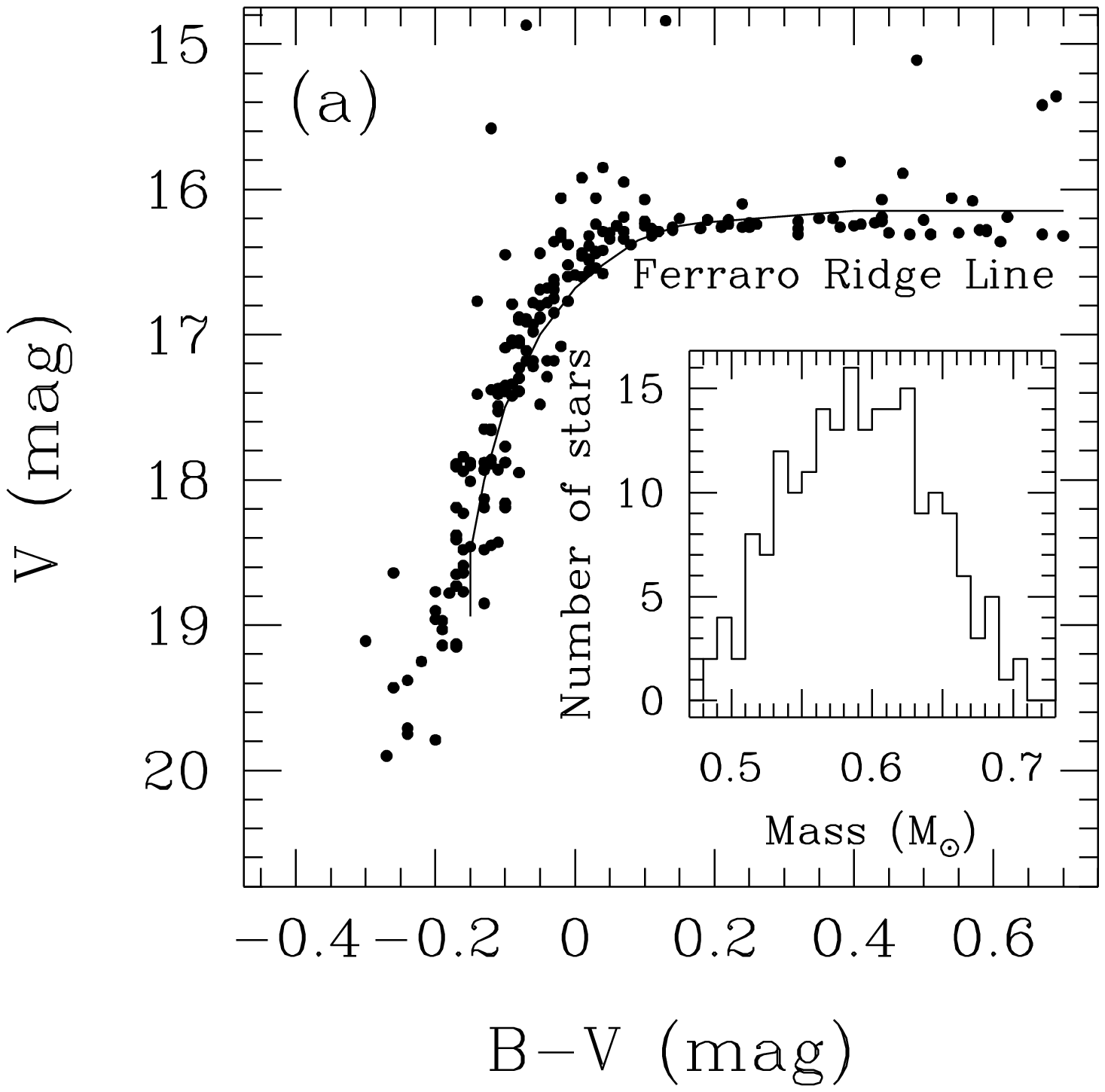}{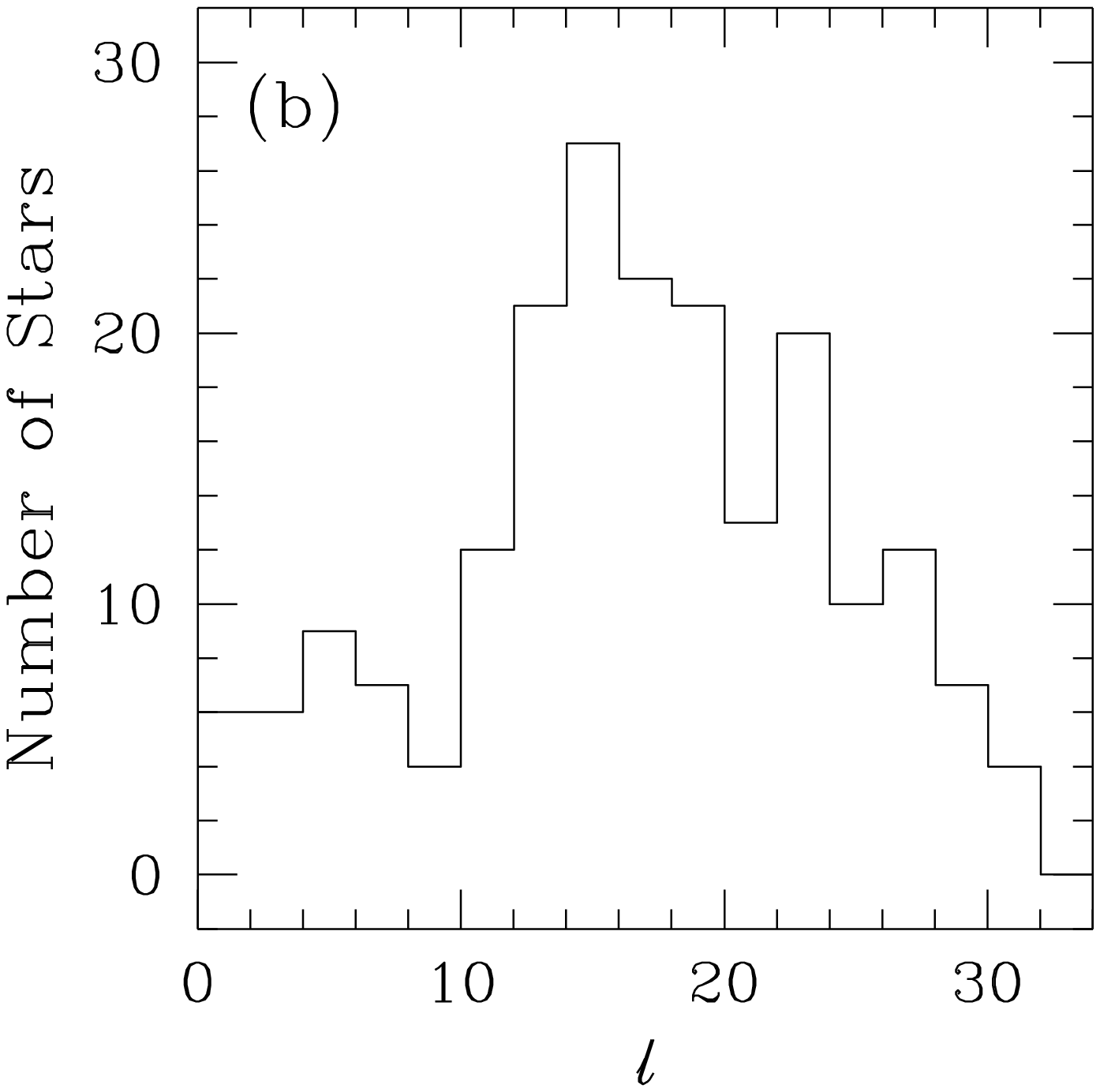}
\caption
{(a) A synthetic color-magnitude diagram (CMD) showing 200 model stars
drawn from the best-fitting Gaussian HB mass distribution. (Stars that
have evolved redward of $\bv = 0.7$ are not shown.) The
observed HB ridge line is plotted for comparison. {\it Inset:} The
distribution of mass among the HB stars in this model.
(b) The HB distribution of the best-fitting Gaussian model.
\label{MODEL} 
}
\end{figure}

We have used the formulation of \shortciteN{LDZ90} in fitting a
Gaussian HB mass-distribution model to the optical CMD of M79
\shortcite{Ferraro92}. This model
assumes that stars on the ZAHB have a
Gaussian distribution in mass $P(M)$ resulting from variable amounts of
mass loss on the RGB:
\begin{equation}
P(M) \propto \left[ M - \left( \langle M_{HB}\rangle - \Delta M \right) \right]
\left( M_{RG} - M \right)
\exp \left\{ - { \left( \frac{ \langle M_{HB}\rangle - M }{\sigma} \right) }^2 \right\} ,
\label {LDZ_dist}
\end{equation}
where $M_{RG} = 0.80 M_{\sun}$ is the mass a star would have at the tip
of the RGB if it did not lose mass (assuming a cluster age of about 15
Gyr), $\Delta M = M_{RG} - \langle M_{HB} \rangle$ is the mean amount
of mass loss, and $\sigma$ is the mass dispersion. The mass
distribution is truncated at $M_{RG}$ at the high-mass end and at
$\langle M_{HB}\rangle - \Delta M$ at the low-mass end (see inset of
\fig{MODEL}a). For most clusters, $\langle M_{HB}\rangle - \Delta M$ is
greater than $M_C$, but
the HB of M79 is so blue that $P(M)$ predicts a finite number of stars
with mass less than $M_C$. These stars, representing only one or two
percent of the total population, are simply ignored in our analysis.

To produce a model HB, 10~000 stars are selected with masses
distributed according to the probability distribution $P(M)$ and ages
chosen from a set of uniform random deviates between 0 and $1.4 \times
10^{8}$ years. We ignore variations in the main-sequence turn-off mass
($< 0.002 M_{\sun}$; \shortciteNP{Castellani80}) over this time scale. Given
the stellar mass and age, we use the HB evolutionary models of
\shortciteN{Dorman93} to determine the temperatures and surface
gravities of the evolving HB and post-HB stars. These models assume
[Fe/H] $= -1.48$, [O/Fe] $= 0.60$, and a core mass of 0.485 $M_{\sun}$.
Stellar magnitudes and colors are then interpolated from the grid of
synthetic stellar flux distributions computed by \shortciteN{Kurucz92} for
stars with an abundance $\mh = -1.5$, assuming a distance 
modulus of $(m - M)_0 = 15.65$ \shortcite{HR79}.

In order to provide a purely empirical measure of the stellar
distribution along the HB, \shortciteN{Ferraro92} define a coordinate ``$l$''
(similar to $X_{HB}$ in the coordinate system of \shortciteNP{RC89}) that is
linear along the ridge line of the HB, divide the whole length into
equal bins, and count the stars populating each bin, perpendicular to
the ridge line (see their Figs. 12 and 13; the technique is described
fully by \shortciteNP{FP93}). We have attempted to implement this process
in software; the resulting distribution is presented in
\fig{M79}b. Note the asymmetric shape: the $l$ distribution is
skewed toward the blue. While noting that each bin in $l$ space may
not correspond to an equal interval in effective temperature,
\shortciteANP{Ferraro92} argue that the stellar distribution along the HB is
not Gaussian.

To determine the values of $\langle M_{HB} \rangle$ and $\sigma$ that
best reproduce the HB of M79, we compute a grid of HB models,
incrementing $\langle M_{HB} \rangle$ and $\sigma$ in steps of 0.01
$M_{\sun}$. Colors and magnitudes of the resulting model HB stars are
scattered according to the observational errors estimated by
\shortciteN{Ferraro92}. For each mass-distribution model, we compute
10~000 model stars, divide them into samples of 200 each, and determine
the $l$ distribution for each sample, ignoring models that have evolved
redward of \bv$ = 0.70$.  We compare the $l$ distribution of each
sample with that of the Ferraro et al. data, using the
Kolmogorov-Smirnov (KS) test \shortcite{PFTV88p558} to determine the
probability $Q_{KS}$ that the model and data are drawn from the same
parent population. By rejecting HB models with $\langle Q_{KS} \rangle <
0.01$, we can exclude models with $\langle M_{HB} \rangle < 0.56$
and $\langle M_{HB} \rangle > 0.61$ at the 99\% confidence level.  With
$\langle Q_{KS} \rangle = 0.37$, the best-fitting model has $ \langle
M_{HB} \rangle = 0.59 M_{\sun} $ and $ \sigma = 0.08 M_{\sun}$.
Because the Gaussian distribution is truncated at both ends, the mass
dispersion parameter in Equation (\ref{LDZ_dist}) is approximately a
factor of two larger than the actual standard deviation of the mass
dispersion, which \shortciteN{LDZ90} call $\sigma_{SD}$. For our model,
$\sigma_{SD} = 0.05 M_{\sun}$.

A synthetic CMD and a plot of the $l$ distribution for 200 HB stars
selected from the best-fitting model are presented in \fig{MODEL}. The
mass distribution of the HB stars in this model is also shown. This
(nearly) Gaussian mass distribution is remarkably successful at
reproducing the non-Gaussian distribution in the parameter $l$ observed
by \shortciteN{Ferraro92}. The transformation from a Gaussian
distribution of initial masses to a skewed distribution of stars along
the HB reflects an evolutionary effect of the \shortciteN{Dorman93} tracks: stars with $M \lesssim 0.50 M_{\sun}$ evolve to higher
luminosities at nearly constant temperature, while those with higher
masses cool rapidly, driving the $l$ distribution sharply toward the red.

The model HB seems to be about 0.2 magnitudes brighter than the cluster
ridge line at $\bv = 0$ and about 0.2 magnitudes fainter at $\bv =
0.4$. These discrepancies may reflect small errors in the evolutionary
or stellar atmosphere models combined with uncertainties in the
cluster's distance modulus. We have repeated the optical CMD fit
assuming a variety of cluster distances and find that the parameters of
the best-fitting model do not change over the range $15.60 \leq ( m - M
)_0 \leq 15.70$.

\section{Fitting the Far-UV Spectrum}

\subsection{The Gaussian Model}
 
To determine whether the Gaussian mass-distribution model that best
fits the optical CMD of M79 is able to reproduce its far-UV
spectrum, we use a Monte Carlo procedure to construct synthetic cluster
spectra. We select stars at random from our best-fitting model HB,
compute their far-UV spectra from the nearest Kurucz (1992) model in
$\log g - T_{eff}$ space, and sum them until the model flux between 900
and 1768 \AA\ equals the HUT flux in this range. By computing
the far-UV spectra for 1000 realizations of this random sampling of the
HB mass distribution, we can accurately assess the probability that the
observed spectrum could have been obtained from the model HB
distribution.

The \shortciteN{Kurucz92} models have a resolution of about 10~\AA,
incorporate statistically correct line strengths for 58 million atomic
and molecular transitions, and are available for scaled solar
abundances $1.0 \gtrsim \mh \gtrsim -5.0$. The models
provide a thorough treatment of line blanketing by metals, though Kurucz's
assumption of local thermodynamic equilibrium (LTE) is inappropriate
for stars with high effective temperatures or low surface gravities.
Fortunately, non-LTE effects are small for stars on the HB
(\shortciteNP{Kudritzki79p295}, \citeyearNP{Kudritzki90}).

Our observed spectrum is binned by 20 pixels to bring it to the 10
\AA\ resolution of the Kurucz models. Model spectra are
fit to the data using the non-linear curve-fitting program ``specfit''
\shortcite{Kriss94}, a $\chi^2$ minimization routine
running in the IRAF\footnote{The Image
Reduction and Analysis Facility (IRAF) is distributed by the National
Optical Astronomy Observatories, which is operated by the Association
of Universities for Research in Astronomy, Inc. (AURA), under
cooperative agreement with the National Science Foundation.}
environment.
The only free parameter in the fit is the normalization of
the model. To model the effects of
interstellar reddening, the program uses a \shortciteN{CCM89} extinction
curve, assuming \ebv\ = 0.01 \shortcite{Peterson93} and
$R_V = 3.1$. A neutral-hydrogen absorption model
assuming a Doppler velocity parameter of 10 km ${\rm s}^{-1}$ is
included in the fit, for which the column density toward M79 is
determined by integrating a 21-cm spectrum from the survey of
\shortciteN{Stark92} through the velocity range $-100$ km ${\rm s}^{-1}$ to
$+100$ km ${\rm s}^{-1}$, with a result of $N_H = 5.307 \times 10^{19}$
cm$^{-2}$. Within specfit, the models are interpolated to
the exact wavelengths of the data points; this resampling may
introduce small errors in the model flux.

In the rebinned spectrum of M79, the signal-to-noise ratio ranges
between 20:1 and 30:1 at wavelengths longer than about 1000 \AA. The
uncertainty in the HUT absolute calibration is estimated to be about
5\% on scales of roughly 50 \AA. Because the systematic errors are
comparable to the random statistical errors, it is necessary to relax
the standard $\chi^2$ criterion for goodness of fit. As a guide to how
well \shortciteN{Kurucz92} models can fit HUT spectra, we consider the
individual globular-cluster UV-bright stars studied by
\shortciteN{DDF94}.  For those stars, the best-fitting models yield
$\chi^2 = 195.0$ for $\nu = 88$ and $\chi^2 = 219.5$ for $\nu = 87$,
respectively. We will thus consider an acceptable spectrum to be one
with $\chi^2 < 200$ and exclude as
unacceptable any ensemble of stars that does not yield $\chi^2 < 200$
in at least 5\% (50 of 1000) of its spectra.
There are 83 data points in the rebinned M79 spectrum.

A UV-bright star discovered with \iue\ by \shortciteN{Altner93} lies at
the edge of the HUT aperture (\fig{UIT}a). The authors estimate the
star to have $T_{eff} =$ 13~000 K, $\log g =$ 3.3, and solar abundance.
Its abundance is important, for though changes in metallicity have only
a minor effect on the flux of Kurucz (1992) models longward of 1200
\AA, line blanketing by metals can vastly alter the apparent continuum
level at shorter wavelengths. \iue\ spectra have a
short-wavelength cutoff of about 1200 \AA, while HUT spectra continue
to the Lyman limit, revealing the extreme sensitivity of the
sub-Ly$\alpha$ continuum to metal-line opacity in stellar atmospheres.
\shortciteANP{DDF94} \citeyear{DDF94,DDF95} fit Kurucz (1992) models to HUT spectra of three UV-bright stars
in globular clusters. In each case, a Kurucz model with the cluster
metallicity provides a significantly better fit to the data than the
best fitting solar-metallicity model.
We thus include a model spectrum of a star with $T_{eff} =$ 13~000 K,
$\log g = 3.3$ (interpolated from $\log g = 3.0$ and 3.5 models), and
$\mh = -1.5$ in each of our synthetic spectra.

We seek to model only the BHB of M79. From its CMD (\fig{M79}a), we see
that the red end of the cluster's BHB lies at $\bv \sim 0.15$,
corresponding to stellar temperatures of about 7500 K.  Because
\shortciteN{Kurucz92} models with $T_{eff} \lesssim 7500$ K do not emit
significant flux at wavelengths shortward of 1770 \AA, and because the
HUT spectrum of M79 rises at longer wavelengths, reflecting the
contribution of stars at the main-sequence turn-off to the integrated
cluster flux, we exclude this region ($\lambda > 1770$ \AA) from our
analysis.

\begin{figure}
\epsscale{0.70}
\plotone{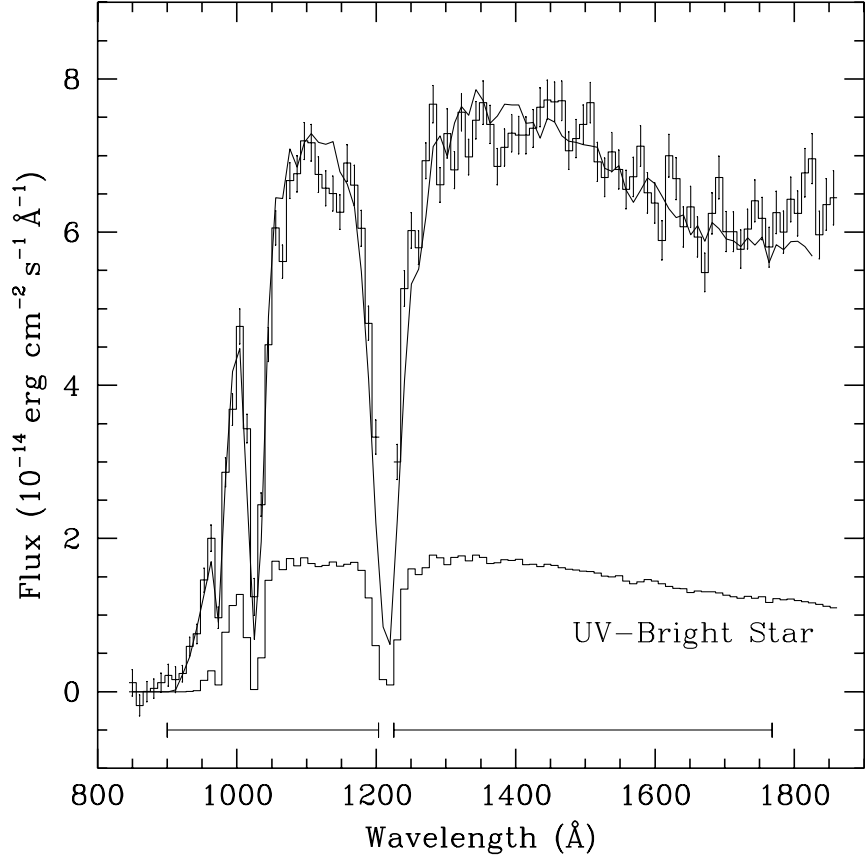}
\caption
{HUT spectrum of M79, with error bars. The data have been airglow
subtracted and flux calibrated, but not dereddened, and are binned by
20 pixels (about 10 \AA). Overplotted is the best-fitting Gaussian
model ($\chi^2 = 307, \nu = 81$) drawn from the Lee \etal\ (1990) mass
distribution that best reproduces the optical CMD of M79 (Ferraro
\etal\ 1992; see Sec.~4.1). The model assumes $\mh = -1.5$
and includes a UV-bright star with $T_{eff} = 13~000$ K, $\log g =
3.3$, whose spectrum is plotted below. The models have been reddened
with a Clayton \etal\ (1989) extinction curve, assuming \ebv\ = 0.01
and $R_V = 3.1$, and include absorption by interstellar hydrogen at a
column density of $N_{H} = 5.307 \times 10^{19} {\rm cm}^{-2}$. Regions
used in the fit are indicated by bars at the bottom of the figure.
\label {HUT_FLUX} 
}
\end{figure}

We construct 1000 synthetic spectra from the 10~000 stars in our model
HB population using the above algorithm and compute $\chi^2$ for each
spectrum. We find that the best-fitting 5\% of the models have values
of $\chi^2 < 426$ ($\nu = 81$) and the best-fitting model has
$\chi^2 = 307$. Its spectrum is presented in \fig{HUT_FLUX}. We see
that the model predicts a \LA\ absorption feature considerably broader
than is shown by the data.  Indeed, most of the contribution to
$\chi^2$ comes from the regions around the \LA\ and $\beta$ absorption
lines. To verify that this discrepancy does not reflect uncertainties
in the \LA\ airglow subtraction, we extract about 1300~s of data from the
darkest part of orbital night, when the \LA\ airglow emission is lowest
and the \LA\ absorption feature least obscured. To these data, we fit a
three-component model, with a linear continuum, a Gaussian absorption
feature, and a model \LA\ airglow profile. The idea is less to
constrain uniquely the relative strengths of the \LA\ absorption and
emission features than to maximize the contribution of the model
\LA\ airglow line. When this model airglow feature is subtracted from
the data, we find no significant change in the resulting absorption
line profile, indicating that the discrepancy between the model and the
data is not due to errors in our airglow subtraction.  We conclude that
the Gaussian mass distribution that best fits the optical CMD of M79 is
unable to reproduce its far-UV spectrum. Hereafter, we shall refer to
this distribution of stars as the Gaussian model.

\subsection{The Empirical Model}

\begin{figure}
\epsscale{0.70}
\plotone{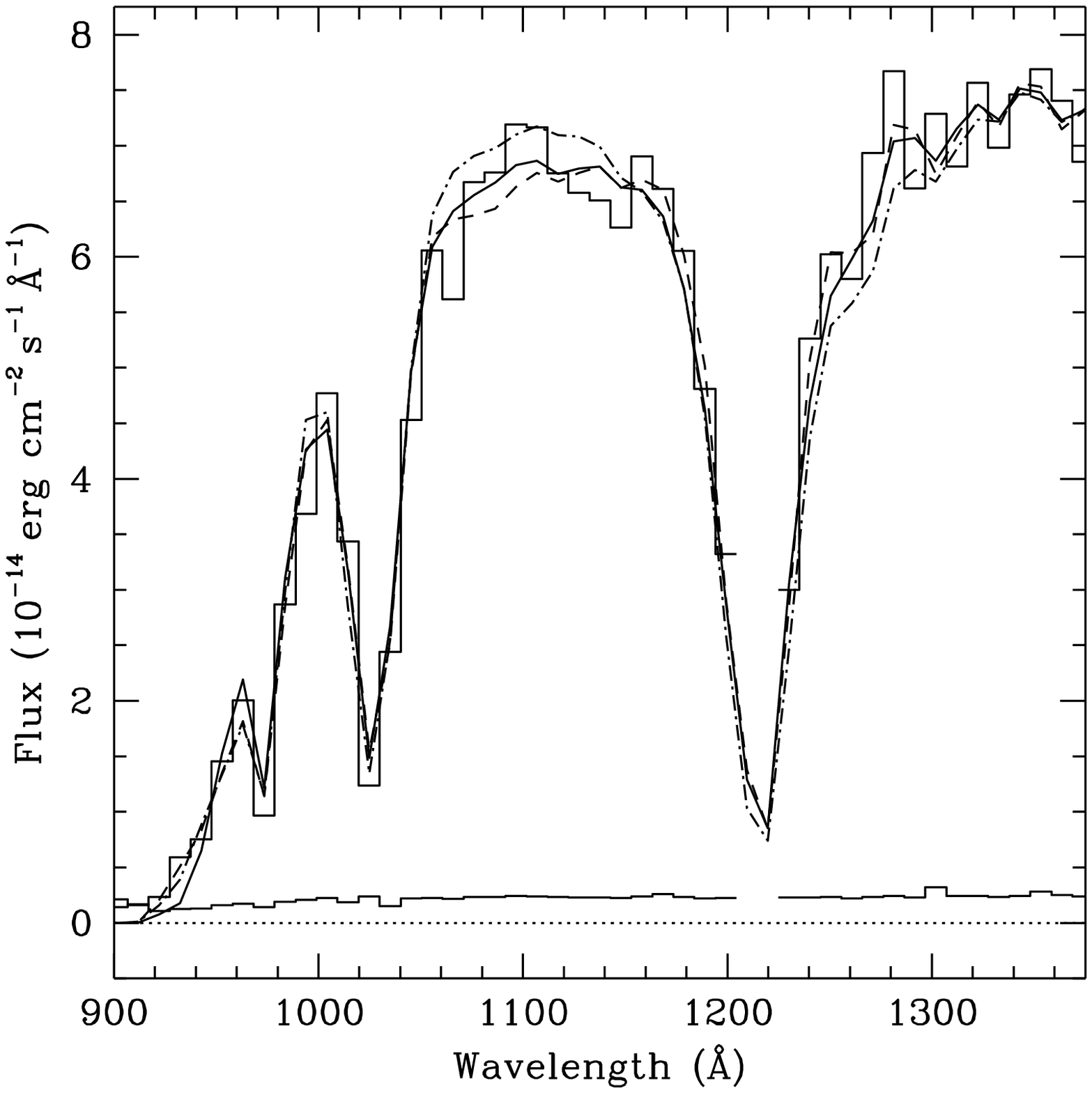}
\caption
{HUT spectrum of M79, reduced as in Fig.~{\protect\ref{HUT_FLUX}}. Overplotted are
three simple models, each composed of 4 to 5 individual Kurucz (1992)
model spectra. The dashed line is the sum of four spectra, all with
$\mh = -1.0$, chosen to minimize $\chi^2$, which equals 148.  The solid
line is composed of four spectra, each with $\mh = -1.5$ but otherwise
unconstrained ($\chi^2 = 171$).  The dot-dashed line represents a
simplified version of our Gaussian model: for each value of $\log g$
between 3.0 and 5.0 (in steps of 0.5 dex), the range of $T_{eff}$ is
limited to that present in the Gaussian model. The model assumes $\mh =
-1.5$ and yields $\chi^2 = 247$. Increases in $\chi^2$ are due
primarily to a broadening of the predicted \LA\ line.  The three models
do not differ significantly longward of 1400~\AA. All are
reddened with a Clayton \etal\ (1989) extinction curve, assuming
\ebv\ = 0.01 and $R_V = 3.1$, and include absorption by interstellar
hydrogen at a column density of $N_{H} = 5.307 \times 10^{19} {\rm
cm}^{-2}$. Error bars for the data are plotted at the bottom of the
figure.
\label {3curves}} 
\end{figure}

We have found that a Gaussian distribution of ZAHB masses is able to
reproduce the CMD of M79, but not its far-UV spectrum.
Is this failure a reflection of errors in our ZAHB mass distribution,
or does it point to more subtle problems in our model? To address this
question, we first attempt to reproduce the cluster's far-UV flux
distribution using a simple collection of \shortciteN{Kurucz92}
stellar atmosphere models. We find that the HUT spectrum of M79 can be
reproduced using only four Kurucz model spectra, all with $\mh
= -1.0$. The models have $T_{eff} =$ 8500, 11~000, 15~000,
and 27~000~K, and $\log g =$ 3.0, 3.5, 4.0, and 3.5, respectively.
Generally speaking, these four spectra provide flux in the regions (1)
longward of about 1500~\AA, (2) between 1500~\AA\ and Ly$\alpha$,
(3) between Ly$\alpha$ and Ly$\gamma$, and (4) shortward of Ly$\gamma$,
respectively. We find that $\chi^2 = 148$ ($\nu = 73$) for this model,
making $\chi^2_{\nu} = 2.0$. The resulting synthetic spectrum is
plotted in \fig{3curves} (dashed line).

We next construct a model stellar population based on this 
distribution of stellar parameters and compare it with the data.
For each HB star observed by \shortciteN{Ferraro92}, we select a
\shortciteN{Kurucz92} model spectrum with $\mh = -1.0$.
We choose $T_{eff}$ to match the observed \bv, then select $\log
g$ according to the following scheme: for stars with $T_{eff} < 9750$ K,
models with $\log g = 3.0$ are selected; for those with $9750 < 
T_{eff} < 13$~000 K, $\log g = 3.5$; for stars with 
13~$000 < T_{eff} < 20$~000 K, $\log g = 4.0$;
and for $T_{eff} > 20$~000 K, $\log g = 3.5$.
The flux is scaled to match the
observed V magnitude. Four stars with measured $\bv < -0.3$ are
excluded from our sample, as their colors are probably erroneous.
Such hot stars would be prominent in the UIT image, yet only one of
them was seen, and its UV color indicates a much lower temperature.

From this ensemble of approximately 200 model stellar
spectra, we generate 1000 synthetic cluster spectra, each including the
\shortciteN{Altner93} UV-bright star (though now with an assumed
metallicity of $\mh = -1.0$), and compare them with our
data. The minimum value of $\chi^2$ that we achieve is 176, and the
lowest 50 values of $\chi^2$ fall below 207. Excluding the UV-bright
star slightly lowers the quality of the resulting fits. We consider
this a successful model and shall refer to it as the ``empirical
model'' in the discussion that follows.

The stars in the empirical model have higher metallicity than the
general M79 population and a somewhat unusual distribution of surface
gravities. Let us investigate these two effects. In \fig{3curves},
we plot the simple, four-component $\mh = -1.0$ model discussed above
(dashed line). Note that it seems to reproduce the \LA\ absorption
feature fairly well. Next we assemble the four Kurucz models with
$\mh = -1.5$ that best reproduce the data. These models
have $T_{eff} = 8500$, 11~000, 15~000, and 29~000 K, and $\log g = 3.5$,
3.5, 3.0, and 3.5, respectively. Combined, the spectra yield $\chi^2 =
170.7$ for $\nu = 74$. The resulting spectrum is plotted as a solid
line in \fig{3curves}; its \LA\ profile appears somewhat wider than 
either our previous model or the data.

\begin{figure}
\plotone{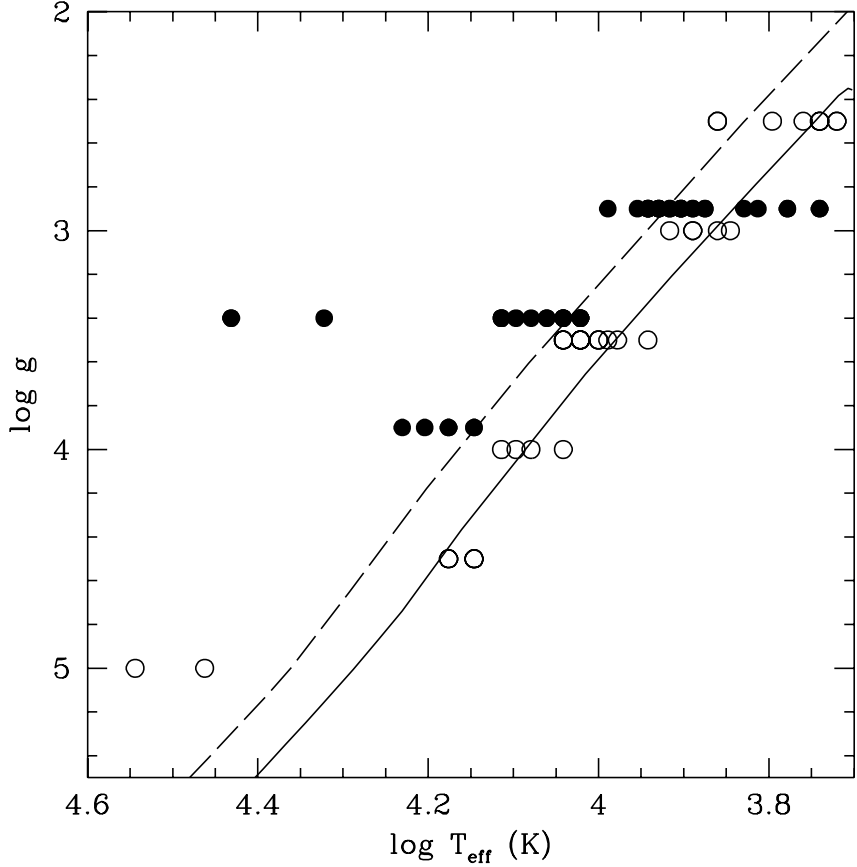}
\caption
{A theoretical $\log g - T_{eff}$ diagram. Plotted as open circles are
the stars from the Gaussian model, based on Equation (1) and
the Dorman \etal\ (1993) evolutionary tracks, that best reproduces the
HUT spectrum ($\chi^2 = 307$). Plotted as solid circles are the stars
from our best-fitting empirical model ($\chi^2 = 176$), based directly
on the cluster's CMD and assuming the $\log g - T_{eff}$ relation
derived in Sec.~4.2. For clarity, the solid circles have been offset by
$-0.1$ dex in $\log g$. Also plotted are the ZAHB (solid line) and the
terminal-age HB (TAHB; dashed line), representing the time of core
helium exhaustion, both from Dorman \etal\ (1993).
\label {gteff} 
}
\end{figure}

Finally, we select individual Kurucz models from our Gaussian model.
\fig{gteff} shows a plot of $\log g$ vs. $T_{eff}$ for the best-fitting
Gaussian and empirical models. The differences between the two models
are discussed below. At each value of $\log g$ between 3.0 and 5.0,
we select values of $T_{eff}$ from the range of temperatures present in
the Gaussian model (open circles).  All of these stars have $\mh
= -1.5$. The best-fitting combination has $T_{eff} = 8250$,
9500, 13~000, 16~000, and 31~000 K, and $\log g = 3.0$, 3.5, 4.0, 4.5,
and 5.0, respectively. $\chi^2 = 247$ for $ \nu = 77$. The model is
shown in \fig{3curves} as a dot-dashed line; its \LA\ absorption
feature is the widest of the three models. These results suggest that
both metallicity and surface gravity may contribute to the poor fit of
our Gaussian model.

One may wonder how such small changes in the \LA\ line width
can produce such large changes in $\chi^2$. Keep in mind that $\chi^2$ is
calculated not from offsets in wavelength but from offsets in flux.
Even small changes in the model line widths imply large changes in
the predicted flux for fixed wavelengths in the wings of the line.
Given our small error bars (shown in \fig{3curves}), $\chi^2$ is quite
sensitive to such discrepancies.

\section{Discussion}

It appears to us more likely that difficulties in fitting the far-UV
spectrum of M79 reflect deficiencies in the model atmospheres than
errors in the HB evolutionary models. In many ways, this problem is
similar to that found by \shortciteN{MHd95}, who analyzed low
and intermediate-resolution optical spectra of stars along the BHB of
M15. They used \shortciteN{Kurucz92} models of both the Balmer line
profiles and the stellar continuum to determine $T_{eff}$, $\log g$,
and $M_*$ and compared these results with the stellar parameters
predicted by \shortciteN{DLV91} for evolved HB stars at the
metallicity of M15 ([Fe/H$] = -2.17$; \shortciteNP{Djorgovski93}).
\shortciteANP{MHd95} found that both the stars' surface gravities
and their masses are systematically lower than is predicted for HB
stars even when luminosity evolution is accounted for.

\fig{gteff} shows a theoretical $\log g - T_{eff}$ diagram. We plot as
open circles the stars from the Gaussian model, based on Equation
(\ref{LDZ_dist}) and the \shortciteN{Dorman93} evolutionary tracks,
that best reproduces the HUT spectrum ($\chi^2 = 307$). Plotted as
solid circles are the stars from our best-fitting empirical model
($\chi^2 = 176$), based directly on the cluster's CMD and assuming the
$\log g - T_{eff}$ relation derived in Sec. 4.2. We see that the stars
of our empirical model have systematically lower values of $\log g$ at
a given effective temperature than do the stars of our Gaussian model.
(The stars with $T_{eff} \lesssim 7500$ K, which do not follow this
trend, are not constrained by our analysis.)
\shortciteN{MHd95} found an offset of about 0.2 dex between the
predicted and measured surface gravities of HB stars in M15.
We also plot in \fig{gteff} the ZAHB (solid line) and the terminal-age
HB (TAHB; dashed line), which represents the time of core helium
exhaustion, both from \shortciteN{Dorman93}. Like
\shortciteANP{MHd95}, we find that stars in our empirical model
tend to lie above the ZAHB and even slightly above the TAHB.

\begin{figure}
\plotone{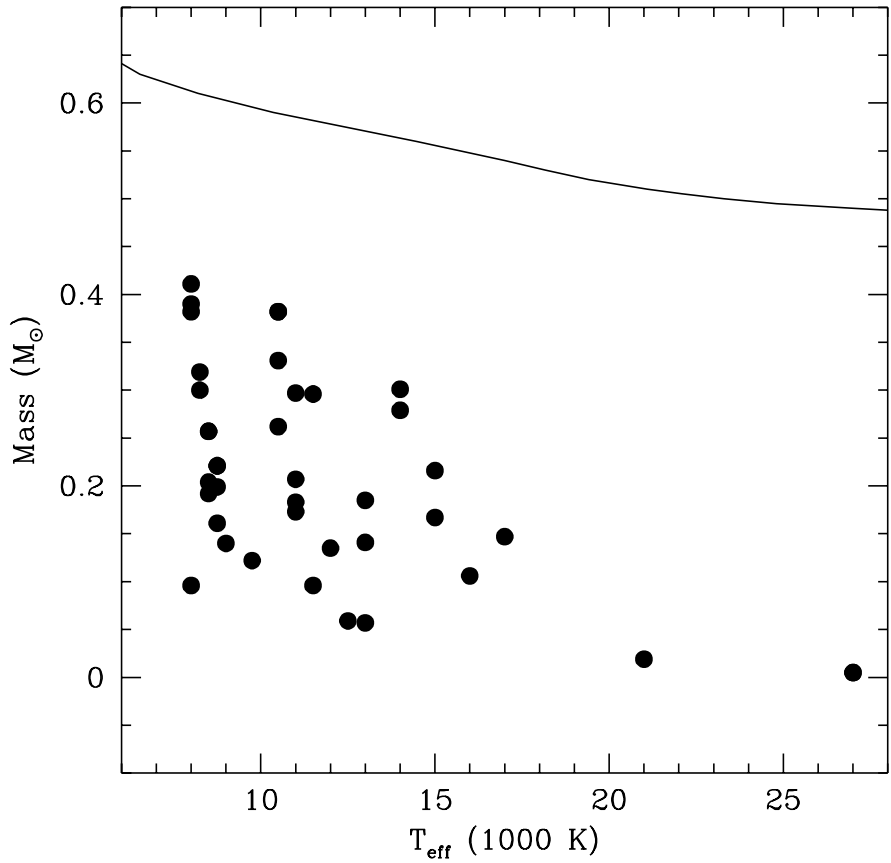}
\caption
{The masses of the stars in our best-fitting empirical model,
scaled to reproduce the $V$ magnitudes of the HB stars observed by
Ferraro \etal\ (1992). The solid line is the ZAHB of Dorman \etal\
(1993). Stars with $T_{eff} < 7500$ K are not included in this plot.
\label {masses} 
}
\end{figure}

\fig{masses} shows the masses of the stars in our best-fitting
empirical model, scaled from the $V$ magnitudes of the cluster's HB
stars. The solid line, again, is the ZAHB of \shortciteN{Dorman93}. Stars with $T_{eff} < 7500$ K are not included in this plot.
The masses calculated for these stars are far lower than is predicted
by standard HB evolutionary theory. \shortciteN{MHd95}
determined the masses of several dozen HB stars in six globular
clusters using published atmospheric parameters. Their masses range
between about 0.2 and 0.5 $M_{\sun}$ at $T_{eff} = $10~000 K, also 
lower than the ZAHB values predicted by \shortciteN{DLV91}.

In presenting Figs.~\ref{gteff} and \ref{masses}, we do not pretend to
have derived accurate surface gravities and masses for the HB stars in
M79 from its optical CMD. Instead, we wish to illustrate how the $\log
g - T_{eff}$ relationship assumed by our empirical model differs from
that predicted by canonical HB evolutionary theory.  The basis for this
relationship remains our model fits to the cluster's far-UV spectrum
and the strong sensitivity of $\chi^2$ to changes in the models'
surface gravity observed in Sec. 4.2. We believe that this result
indicates a real discrepancy between HB evolutionary theory and the
models most commonly used to determine stellar parameters, if not
between the theory and the stars themselves.

The high metallicity of the stars in the empirical model may be an
artifact of the uncertainties in their surface gravities, or may
reflect genuine abundance anomalies in the cluster's HB stars. In
Kurucz's (1992) model spectra, the apparent far-UV continuum level
is determined by the combined absorption of innumerable metal lines.
The completeness of the
absorption-line models and the relative abundances of the absorbing
species determine the shape of the resulting model spectrum.  The
models assume scaled solar abundances, but this is certainly
inappropriate. Giant-branch stars in 47 Tuc ([Fe/H$] = -0.71$;
\shortciteNP{Djorgovski93}), for example, have non-solar abundance
ratios, with enhancements in aluminum and $\alpha$ elements
\shortcite{BWO90,BW92}. These abundance anomalies contribute line
opacity that is not included in the Kurucz models.

In both the work of \shortciteN{MHd95} and in this paper,
discrepancies in the hydrogen line widths are responsible for the poor
fits at the surface gravities predicted by the evolutionary models: in
their work it is the Balmer series, whereas in ours it is the Lyman
series. We suggest that improving the treatment of hydrogen absorption
in the \citeN{Kurucz92} spectral synthesis codes may help to solve
this problem. This is an important task, as it would allow the joint
optical/UV technique we have adopted to provide meaningful constraints
on the mass distribution of the HB, and the Balmer series line fits to
provide accurate masses for individual HB stars.  Given the current
discrepancies between derived stellar parameters and HB evolutionary
theory, however, we conclude that it is not yet possible to combine
these models to provide meaningful constraints on the ZAHB mass
distribution in globular clusters.

We {\it can} place limits on the temperature distribution of the HB
stars in M79. UIT set an upper limit of 25~000 K to the temperature
of stars farther than 1\arcmin\ from the cluster center \shortcite{UIT_M79}. The \shortciteN{Ferraro92} optical CMD contains four stars
with $-0.3 < \bv < -0.25$, indicating 27~000~K $\lesssim T_{eff}
\lesssim$ 32~000~K. While these particular measurements may be
uncertain, we find that each of the ten best-fitting synthetic spectra
from our empirical model includes a substantial contribution from stars
with $T_{eff} \gtrsim$ 27~000~K. For example, the simple,
four-component $\mh = -1.0$ model, plotted as a dashed line in
\fig{3curves}, includes a flux contribution from Kurucz models with
$T_{eff} = 27~000$~K corresponding to 2.2 ZAHB stars.  Only such stars can
produce the observed flux shortward of Ly$\gamma$; thus, at least a few
hot stars must reside within about 1\arcmin\ of the cluster center. We
see no evidence, however, of a large population of previously unobserved
extreme HB stars---or of any objects with $T_{eff} \gtrsim$ 32~000~K.

\section{Conclusions}

We have shown that the apparently asymmetric distribution of HB stars
in M79 observed by \shortciteN{Ferraro92} can be reproduced with a
ZAHB mass-distribution function in the form of a Gaussian with
polynomial truncation terms if stellar evolution is properly taken into
account.  The function that best fits the HB of M79 has a mean mass of
$0.59 M_{\sun}$ and a standard deviation of $0.05 M_{\sun}$. This study
represents one of the first detailed investigations into the underlying
stellar mass distribution on the HB.

The synthetic HB distribution that best fits the optical CMD is,
however, unable to reproduce the HUT far-UV spectrum of the cluster.
\shortciteN{Kurucz92} models fit directly to the HUT spectrum have
values of $\log g$---and thus stellar masses---significantly lower than
are predicted by canonical HB evolutionary theory. This result is
consistent with the recent report by \shortciteN{MHd95} that
\shortciteN{Kurucz92} stellar atmosphere models yield surface
gravities for BHB stars systematically lower than are predicted by the
evolutionary tracks of \shortciteN{DLV91}. \shortciteANP{Kurucz92} models with $\mh = -1.0$ provide significantly better fits to the
cluster spectrum than do those with $\mh = -1.5$. These discrepancies
may reflect non-solar abundances ratios in the HB stars of M79 and/or
problems in the treatment of hydrogen absorption by the Kurucz model
atmospheres. The deficiencies are made obvious here by the high
signal-to-noise ratio of the HUT spectra and the reliability of their
calibration.

Stars with 27~000 K $\lesssim T_{eff} \lesssim$ 32~000 K are required
to reproduce the observed flux shortward of Ly$\gamma$. Thus, at least
a few such objects must reside within about 1\arcmin\ of the cluster
center, and should be detectable using WFPC2 on \hst. We see no
evidence, however, of a large population of previously unobserved
extreme HB stars.

\acknowledgments

We would like to thank R. Rood for helpful discussions; W. Landsman for
kindly providing a UIT image of M79 (FUV0141NC) and UIT coordinates
and magnitudes for the cluster's HB stars; L. Danly and K. Kuntz for
providing a 21-cm spectrum taken in the direction of M79
\cite{Stark92}; and R. Kurucz for providing a computer-readable tape of
his stellar atmosphere models. This research has made use of the Simbad
database, operated at CDS, Strasbourg, France. We wish to acknowledge
the efforts of our colleagues on the HUT team as well as the NASA
support personnel who helped make the Astro-1 mission successful. The
Hopkins Ultraviolet Telescope Project is supported by NASA contract
NAS5-27000 to the Johns Hopkins University. BD acknowledges support
from NASA RTOP 188-41-51-03.

\newpage


\newpage

\end{document}